\documentclass[conference]{IEEEtran}
\IEEEoverridecommandlockouts

\usepackage{cite}
\usepackage{amsmath,amssymb,amsfonts}
\usepackage{algorithmic}
\usepackage{graphicx}
\usepackage{textcomp}
\usepackage{xcolor}
\usepackage{url, booktabs, multirow}

\def\BibTeX{{\rm B\kern-.05em{\sc i\kern-.025em b}\kern-.08em
    T\kern-.1667em\lower.7ex\hbox{E}\kern-.125emX}}
\begin{document}

\title{Dense-Sparse Dynamic Time Warping for Customizing Piano Concerto Accompaniments\\
}

\author{\IEEEauthorblockN{TJ Tsai\thanks{\copyright~2025 IEEE. Personal use of this material is permitted. Permission from IEEE must be obtained for all other uses, in any current or future media, including reprinting/republishing this material for advertising or promotional purposes, creating new collective works, for resale or redistribution to servers or lists, or reuse of any copyrighted component of this work in other works.}}
	\IEEEauthorblockA{
		\textit{Harvey Mudd College}\\
		Claremont, CA USA\\
		ttsai@g.hmc.edu}
	\and
	\IEEEauthorblockN{Kavi Dey}
	\IEEEauthorblockA{
		\textit{Harvey Mudd College}\\
		Claremont, CA USA\\
		kdey@g.hmc.edu}
	\and
	\IEEEauthorblockN{Yigitcan \"{O}zer}
	\IEEEauthorblockA{
		\textit{International Audio Laboratories}\\
		Erlangen, Germany\\
		yigitcan.oezer@audiolabs-erlangen.de}
	\and
	\IEEEauthorblockN{Meinard M\"{u}ller}
	\IEEEauthorblockA{
		\textit{International Audio Laboratories}\\
		Erlangen, Germany\\
		meinard.mueller@audiolabs-erlangen.de}
}
\maketitle

\begin{abstract}
In this study, we explore how pianists can customize Music Minus One (MMO) concerto accompaniments to match their playing style. Bypassing the need for a symbolic score, often not available digitally, we use three types of audio data: solo piano recordings, MMO orchestra-only recordings, and mixed recordings of both piano and orchestra (e.g., from YouTube). The mixed recording serves as an intermediary reference to align the solo and orchestra parts, with only the orchestral part being adjusted through time-scale modification to synchronize with the user's playing.  The main challenge with estimating these alignments is the spectral mismatch between recordings containing different musical parts.  Motivated by this application scenario, we introduce Dense-Sparse DTW, a variant of Dynamic Time Warping (DTW) that is designed to improve robustness of alignments to spectral mismatch by focusing on aligning a selected subset of audio frames containing prominent timing cues.  We collect and annotate data from four piano concerto movements and establish a framework for generating and evaluating customized accompaniment recordings.  On this benchmark, we show that Dense-Sparse DTW has better or comparable performance than more complex approaches based on source separation and spectral subtraction techniques.
\end{abstract}


\section{Introduction}
\label{sec:introduction}

This paper explores an application in which a pianist can modify a Music Minus One recording to match their playing style.  Music Minus One (MMO) is a company that sells recordings of the accompaniment part (only) of classical, jazz, and popular music, so that a musician can play their part with accompaniment.  One of the drawbacks of MMO is that the musician has no control over the accompaniment recording.  In this work, we explore an offline accompaniment generation task in which an existing MMO recording is modified to match the user's playing.

Automatic musical accompaniment has a rich history of study that extends several decades (e.g. see \cite{hidaka1995automatic, raphael2001probabilistic, dannenberg2006music, raphael2010music, xia2017improvised, pasini2024bass} for representative samples).  For non-improvised music, previous work generally assumes that the accompaniment system has access to a musical score in a symbolic format like MIDI.  However, for certain genres of music (such as piano concertos), symbolic scores are not widely available and are very time-consuming to create.  

In this work, we explore a paradigm for automated piano concerto accompaniment in which a symbolic score is not needed.  Instead, we assume that we only have access to three forms of audio data: (1) recordings of the user’s piano playing without any orchestral accompaniment (which we refer to as the piano-only recording or $P_\mathrm{user}$), (2) an MMO recording containing the orchestral accompaniment part only (which we refer to as the orchestra-only recording or $O_\mathrm{acc}$), and (3) a complete recording containing both piano and orchestra such as one might find on Youtube (which we refer to as the mix recording or $M_\mathrm{ref}$).  $M_\mathrm{ref}$ serves as an intermediary reference to facilitate the alignment between $P_\mathrm{user}$ and $O_\mathrm{acc}$.  In a sense, $M_\mathrm{ref}$ takes over the role of the ``score'' which specifies both the piano and orchestra parts.  Given this formulation, Figure 1 shows a straightforward blueprint for customizing an MMO recording: estimate the $P_\mathrm{user}$--$M_\mathrm{ref}$ and $M_\mathrm{ref}$--$O_\mathrm{acc}$ alignments, use these estimates to infer the $P_\mathrm{user}$--$O_\mathrm{acc}$ alignment, and then time-scale modify $O_\mathrm{acc}$ to produce $O_\mathrm{user}$ that is synchronous to $P_\mathrm{user}$.  

The biggest challenge in estimating these alignments is the spectral mismatch between $M_\mathrm{ref}$, $P_\mathrm{user}$, and $O_\mathrm{acc}$ (since each contains different musical parts).  One could approach this problem in a number of ways using existing techniques: (a) simply ignore the spectral mismatch and estimate the alignments using standard dynamic time warping (DTW), (b) apply source separation to $M_\mathrm{ref}$ to estimate the piano and orchestra components (e.g.~\cite{ozer2022source, OezerM24_PianoSourceSep_TASLP}), and then align the estimated components against $P_\mathrm{user}$ and $O_\mathrm{acc}$, or (c) align the dominant soloist part ($P_\mathrm{user}$) against $M_\mathrm{ref}$, subtract its spectral components from $M_\mathrm{ref}$'s spectrogram, and then align the modified spectrogram against $O_\mathrm{acc}$ (e.g.~\cite{yang2021aligning}).  In the process of trying these approaches, we discovered a surprisingly simple yet effective approach.

The key insight is that by focusing only on aligning selected audio frames that contain prominent timing cues, we can reduce the impact of spectral mismatch on the estimated alignment.  For example, the orchestra part often contains rests or silences in which the soloist plays alone.  If one tries to align these regions of silence in $O_\mathrm{acc}$ against $M_\mathrm{ref}$, the spectral mismatch will lead to noisy alignments.  If, however, we ignore temporal regions (like silence) in $O_\mathrm{acc}$ that lack prominent timing cues, we can focus on aligning the more ``robust'' parts and simply use linear interpolation in the less ``robust'' regions.  We propose a variant of DTW called Dense-Sparse DTW (DS-DTW) that aligns selected elements from one sequence (the ``sparsified'' sequence) against another (dense) sequence.

DS-DTW fits into a broad corpus of work on DTW as a tool for alignment.  Works in this area generally fall into one of two groups.  The first group focuses on reducing the quadratic computation and memory costs of DTW.  Some works speed up exact DTW through the use of lower bounds \cite{zhang2011inner}, early abandoning \cite{rakthanmanon2012searching}, or specialized hardware \cite{wang2013accelerating}.  Other works approximate DTW by imposing allowable bands in the cost matrix \cite{sakoe1978dynamic, itakura1975minimum}, estimating alignments at multiple resolutions \cite{MuellerMK06_EfficientMultiscaleApproach_ISMIR, SalvadorC04_fastDTW}, or working with a fixed amount of memory \cite{PraetzlichDM16_MsDTW_ICASSP}.  For sparse time series data, several methods have been proposed to speed up exact or approximate DTW \cite{hwang2022fast, mueen2018speeding}.  The second group focuses on extending the behavior of DTW in some way.  Examples in music processing include handling repeats and jumps \cite{Grachten2013AutomaticAO, shan2020improved}, performing alignment in an online setting \cite{dixon2005live, macrae2010accurate}, handling partial alignments \cite{MuellerA08_PathConstrained_ICASSP}, using multiple performances to improve alignment accuracy \cite{wang2016robust}, and handling pitch drift in a capella music \cite{waloschek2018driftin}.  DS-DTW falls into the second group, with an emphasis on improving the robustness of alignments to spectral mismatch.


\begin{figure}
	\centerline{\includegraphics[width=\columnwidth]{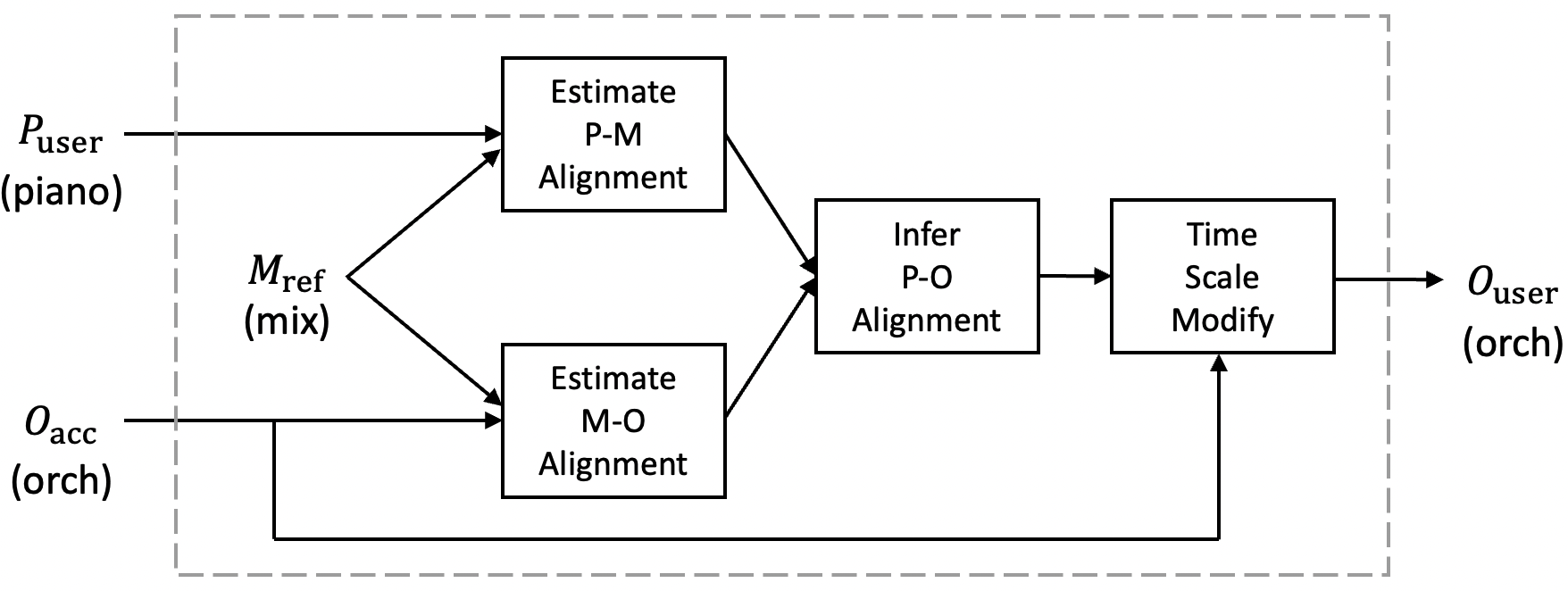}}
	\caption{Overview of a system for customizing concerto accompaniments.  $P_\mathrm{user}$ is the user's piano playing, $O_\mathrm{acc}$ is a Music Minus One recording of the orchestra part, $M_\mathrm{ref}$ is a mix recording (e.g.,~from Youtube), and $O_\mathrm{user}$ is the customized orchestral accompaniment.}
	\label{fig:SystemOverview}
\end{figure}

This paper has two main contributions.  First, we collect a set of data that allows for systematic study of a score-free paradigm for automated piano concerto accompaniment (Section~\ref{sec:benchmark_design}).  This dataset involved selecting MMO and mix recordings, recruiting pianists to record performances of selected concerto movements, and annotating the timestamps of measure downbeats in the collected recordings.  Using this data, we designed a benchmark framework for assessing the alignment accuracy of an adaptive concerto accompaniment system.  Our framework is designed to allow for systematic study of both offline and online concerto accompaniment generation, though in this initial study we focus only on the offline task.  Our data, benchmark framework, and code are all open source except for the MMO recordings.\footnote{\url{https://github.com/HMC-MIR/PianoConcertoAccompaniment}}  As our main technical contribution, we introduce a novel offline alignment algorithm called Dense-Sparse DTW, which is designed to improve the robustness of alignments to spectral mismatch by focusing on selected audio frames (Section~\ref{sec:systemDescr}).  We show that this algorithm leads to comparable or better performance than other systems based on standard alignment, source separation, and spectral subtraction techniques (Section~\ref{sec:results}).

\section{Benchmark Design}
\label{sec:benchmark_design}

In this section we describe the audio and score data, data pre-processing, and evaluation methodology in our benchmark.

\subsection{Audio Data}
\label{subsec:audio_data}

The benchmark contains three types of audio: orchestra-only ($O_\mathrm{acc}$), piano-only ($P_\mathrm{user}$), and mix ($M_\mathrm{ref}$) recordings.  

\textit{$O_\mathrm{acc}$ recordings}.  The orchestra recordings consist of four concerto movements from the Music Minus One catalog shown in Table \ref{tab:overview}.  Our entire framework is open source except for these orchestra recordings.  We annotated the $O_\mathrm{acc}$ recordings with timestamps of measure downbeats.  These timestamps serve as reference instances in order to evaluate alignment accuracy.

\textit{$P_\mathrm{user}$ recordings}.  The piano-only recordings were collected by recruiting pianists to perform the concerto movements (one pianist per movement).  The pianists were asked to play along to the $O_\mathrm{acc}$ recordings, so that their playing would be synchronized.  The pianists listened to the accompaniment tracks on headphones while performing and recording the soloist part on an acoustic piano.\footnote{Note that, even though the $P_\mathrm{user}$ and $O_\mathrm{acc}$ recordings are already synchronized by design, they are never directly aligned -- we compute the $P_\mathrm{user}$--$M_\mathrm{ref}$ and $O_\mathrm{acc}$--$M_\mathrm{ref}$ alignments (which are nontrivial) to infer the $P_\mathrm{user}$--$O_\mathrm{acc}$ alignment.  We chose to record the piano parts in this way to enable us to study other formulations involving source separation.}

\textit{$M_\mathrm{ref}$ recordings}.  The mix recordings are taken from the IMSLP website.{\footnote{\url{https://imslp.org/}}  We found two recordings for each concerto, which ranged in quality from historic performances to performances by university students.  Because measure annotations are time-consuming to create, we only annotated measure downbeats for one selected mix recording from each concerto movement.  These annotations help diagnose how much of the $P_\mathrm{user}$--$O_\mathrm{acc}$ alignment error is coming from the $P_\mathrm{user}$--$M_\mathrm{ref}$ alignment versus the $M_\mathrm{ref}$--$O_\mathrm{acc}$ alignment.
	
Table \ref{tab:overview} summarizes the audio data in our benchmark, which we call the Concerto Accompaniment Benchmark.
We note that this data collection is distinct from the recently released Piano Concerto Dataset (PCD)~\cite{OezerSAJEM_PCD_TISMIR}.  The PCD consists of short 12-second excerpts to facilitate the study of piano--orchestra source separation, whereas our data set consists of entire movements to facilitate the study of alignment and accompaniment generation.
	
\begin{table}
	\caption{Overview of the audio data in the Concerto Accompaniment Benchmark.}
	\centering
		\begin{tabular}{lcccc}
			\toprule
			Concerto Movement & \#P & \#O & \#M &  Total Dur \\
			\midrule
			Rachmaninov No 2, Op 18, Mov 1 & $1$ & $1$ & $2$ &  0:43:32 \\
			Mozart No 21, K467, Mov 1 & $1$ & $1$ & $2$ & 0:52:42 \\
			Beethoven No 1, Op 15, Mov 1 & $1$ & $1$ & $2$ & 1:01:30 \\
			Bach No 5, BWV 1056, Mov 1 & $1$ & $1$ & $2$ & 0:13:21 \\
			\midrule
			Total & $4$ & $4$ & $8$ & 2:51:05 \\
			\bottomrule
		\end{tabular}
	\label{tab:overview}
\end{table}
	
\subsection{Score Data}
\label{subsec:score_data}

The benchmark requires score annotations for evaluation purposes (only).  Since extended silence has an ambiguous ground truth alignment, we only evaluate alignment accuracy when both piano and orchestra are active.  These considerations are described below.

\textit{Terminology}.  The sheet music score consists of a sequence of measures, where each measure is further subdivided into beats.  We define a \emph{piano run} as a contiguous sequence of beats in which the piano part is active.  Because the piano part has rests of various duration, we consider the piano part to be active as long as rests are two measures or less in duration.  We can partition the beats in which the piano part is active into a set of non-overlapping piano runs of maximal duration, which we call \textit{piano chunks}.

\textit{Score annotations}.  For each concerto movement, we found one sheet music PDF from IMSLP and annotated it in two ways.  First, we partition the score into piano chunks and annotate the corresponding measure numbers.  The number of piano chunks in the Rachmaninoff, Mozart, Beethoven, and Bach movements was 4, 5, 5, and 1, respectively.  Second, we identify the measures which should be used to evaluate alignment accuracy.  We only evaluate alignment accuracy on measures where both piano and orchestra are active.

\subsection{Data Pre-Processing}
\label{subsec:data_preprocessing}

Several additional preprocessing steps were carried out to prepare the data for experimentation.

\textit{Query Generation}.  We define a query $P_\mathrm{query}$ to be an audio recording of the user playing a single piano chunk.  In Figure \ref{fig:SystemOverview}, the query is the input to the accompaniment generation system.  Because we had limited data, we performed data augmentation by considering time-scale modified versions of each query.  Time-scale modification (TSM) is a technique for changing the tempo of a musical recording without changing its pitch.  We performed TSM using the method described in~\cite{DriedgerME14_HPTSM_IEEE-SPL} based on harmonic-percussive source separation.  We used the set of logarithmically symmetric TSM factors $\{0.8, 0.9, 1, 1.11, 1.25\}$.  

\textit{Scenario Generation}.  We define a scenario to be a tuple of three audio recordings: a piano only $P_\mathrm{user} = P_\mathrm{query}$ recording of a single piano chunk, an orchestra only recording $O_\mathrm{acc}$ of the entire piece, and a mix recording $M_\mathrm{ref}$ of the entire piece.  We consider all possible combinations of $P_\mathrm{query}$, $O_\mathrm{acc}$, and $M_\mathrm{ref}$ recordings.  In total, there are $150$ scenarios in the benchmark.

\subsection{Evaluation}
\label{subsec:evaluation}

The quality of the generated concerto accompaniment recording depends on both objective and subjective factors.  In our initial framework, we focus on the part of system performance that can be measured objectively: alignment accuracy.  We measure this by comparing the estimated $P_\mathrm{query}$--$O_\mathrm{acc}$ alignment against the annotated downbeat timestamps.  At each annotated downbeat in $P_\mathrm{query}$, we calculate the alignment error between the estimated and annotated locations in $O_\mathrm{acc}$.  We characterize system performance on each concerto movement as an error rate indicating the percentage of alignment errors larger than a fixed error tolerance, where we consider a range of error tolerances.  We then aggregate system performance across the benchmark by averaging the error rates on each movement.\footnote{This weights each concerto movement equally.  Some movements are much longer than others (e.g., 16 min for Beethoven vs 3.5 min for Bach), so this ensures that no one movement dominates the benchmark due to its duration.}

\section{System Description}
\label{sec:systemDescr}

In this section we give an overview of our proposed approach and describe the Dense-Sparse DTW alignment algorithm.

\subsection{Overview}
\label{subsec:overview}

The overall approach to aligning piano-only ($P_\mathrm{query}$) and orchestra-only recordings ($O_\mathrm{acc}$) is to use a mix recording ($M_\mathrm{ref}$) (containing both piano and orchestra parts) as an intermediary representation.  There are three steps to this process.  First, we align $P_\mathrm{query}$ with $M_\mathrm{ref}$ using subsequence DTW with chroma features and cosine distance metric.  Subsequence DTW is a variant of DTW that finds the optimal alignment between a short query sequence and any subsequence within a reference sequence \cite{muller2021fundamentals}.  We allow for step size transitions of \{(1,1), (1,2), (2,1)\} with multiplicative transition weights 1, 1, 2, respectively, which assumes that the tempos in the recordings differ by a factor of two at most.  By using the piano-only chunk as the query, we can identify the matching region of the mix recording.  Even though there is a spectral mismatch between $P_\mathrm{query}$ and $M_\mathrm{ref}$, we find experimentally that this simple method yields fairly reliable alignments due to the fact that the soloist piano part is dominant in the mix recording and contains frequent alignment cues (like note onsets).  Second, we align $O_\mathrm{acc}$ with $M_\mathrm{ref}$ using Dense-Sparse Dynamic Time Warping (DS-DTW).  Unlike the piano part, the orchestral part is not dominant in the mix recording and often contains rests and silences, which results in a much more severe spectral mismatch between $O_\mathrm{acc}$ and $M_\mathrm{ref}$.  This motivates the need for an alignment approach like DS-DTW that is more robust in such scenarios.  Third, we use the estimated $P_\mathrm{query}$--$M_\mathrm{ref}$ and $M_\mathrm{ref}$--$O_\mathrm{acc}$ alignments to infer the $P_\mathrm{query}$--$O_\mathrm{acc}$ alignment.

\subsection{Dense-Sparse DTW}
\label{subsec:sparsedtw}

DS-DTW estimates the alignment between a dense sequence and selected elements from another sequence that contain useful alignment cues (the ``sparsified'' sequence).  In our application, we use DS-DTW to align the mix features ($m_1$,$\dots$, $m_N$) and selected elements from the orchestra features (a subset of $o_1$,$\dots$, $o_K$).  The algorithm has four main steps.

The first step is feature selection.  This is done in three substeps.  Given the orchestra features $o_1$, $o_2$, $\dots$, $o_K$ where $o_t \in \mathbb{R}^d$ (in our case, chroma features), the first step is to compute the flux $f_t \in \mathbb{R}$ at each time index $t$, calculated as $f_t = || o_t - o_{t+1} ||_1$.  The second step is to select a threshold $\tau$ for deciding which features to keep.  The intuition here is that if the feature vector is not changing, there is no benefit in performing a dense alignment.  To select the threshold, we define the desired fraction $\gamma$ of features to keep, and then select $\tau$ to ensure that $L \approx \gamma \cdot K$ feature vectors have flux values higher than $\tau$.  The third step is to select all features $o_{t_1}$, $o_{t_2}$, $\dots$, $o_{t_L}$ whose flux values satisfy $f_{t_i} > \tau$, $i\in\{1,\dots,L\}$ where $1 \le t_1 < t_2 < \dots < t_L \le K$.  In addition, we also compute the gap lengths between adjacent selected feature vectors $g_i = t_{i+1} - t_i$, $i \in \{1,\dots,L-1\}$.

The second step is to compute the pairwise cost matrix $C \in \mathbb{R}^{L \times N}$ between $o_{t_1}$,$\dots$, $o_{t_L}$ and $m_1$,$\dots$, $m_N$.  This is done using a cosine distance metric.  Note that this matrix indicates pairwise costs between the \textit{selected} orchestra features and all of the mix features.

\begin{figure}
	\centerline{\includegraphics[width=\columnwidth]{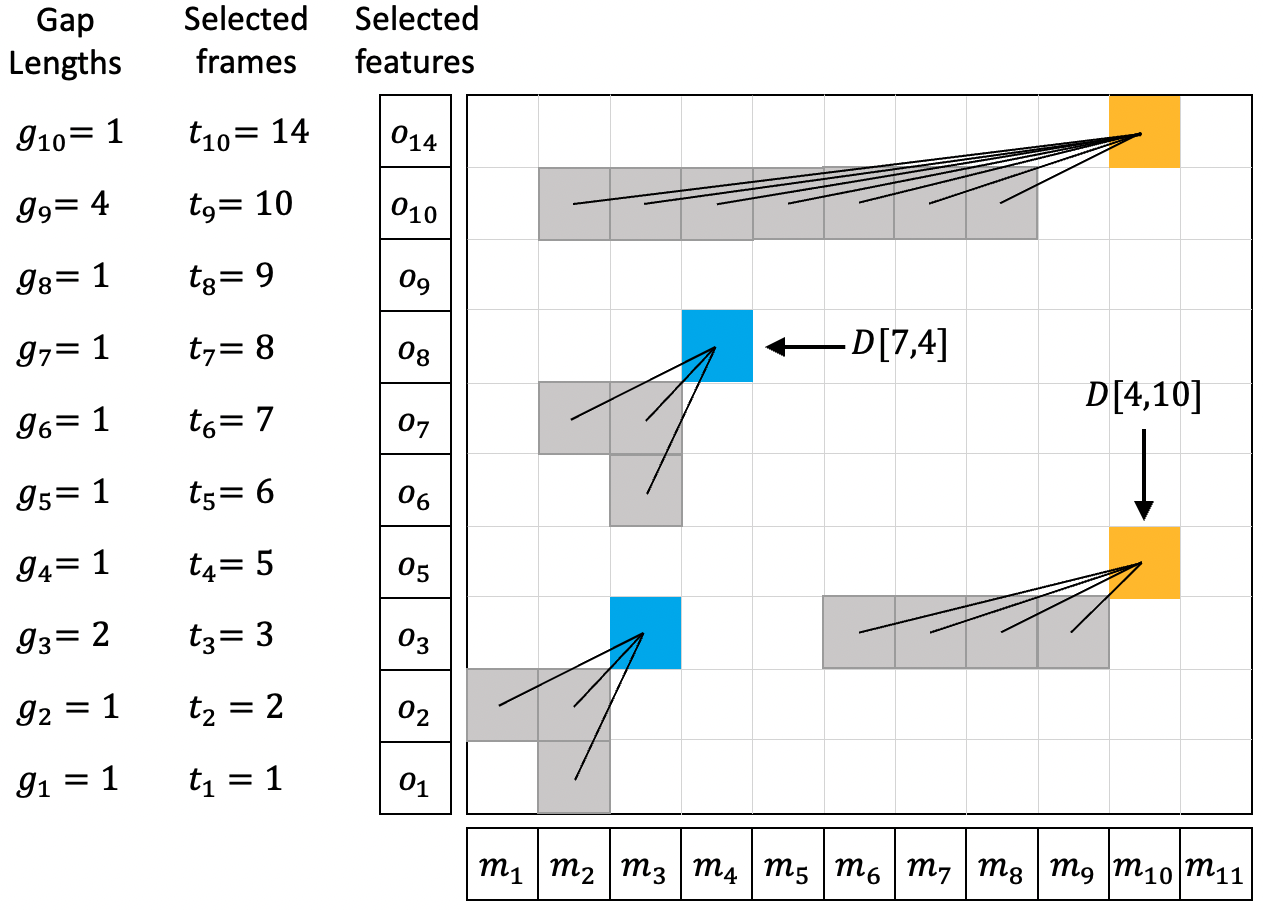}}
	\caption{Visualization of DS-DTW for aligning a sparse sequence (selected orchestra chroma features, vertical axis) against a dense sequence (mix chroma features, horizontal axis).  During dynamic programming, there are two different modes of operation: dense matching (blue) and sparse matching (orange).}
	\label{fig:DualModeExpl}
\end{figure}

The third step is to compute a cumulative cost matrix $D \in \mathbb{R}^{L \times N}$ and backtrace matrix $B \in \mathbb{Z}^{L \times N}$ using dynamic programming.  We initialize $D[1,:] = C[1,:]$, which allows the alignment path to begin anywhere in the mix recording.  We also initialize $D[i,1]=\infty$, $i>1$.  As with standard DTW, $D[i,j]$ indicates the cumulative path cost of the optimal path up to and including position $(i,j)$.  However, unlike standard DTW, DS-DTW has two different modes of operation: one mode for handling dense alignments (blue squares in Figure \ref{fig:DualModeExpl}) and one mode for handling gaps (orange squares in Figure \ref{fig:DualModeExpl}).  To determine which mode to use when calculating $D[i,j]$, we simply check the gap lengths of the previous two selected features, i.e.~$g_{i-1}$ and $g_{i-2}$.  If $g_{i-1} = 1$ and $g_{i-2} = 1$, then we select the dense matching mode and calculate $D[i,j]$, $i>1$, $j>1$ using the standard (subsequence) DTW formulation:\footnote{In both equation \eqref{eqn:denseMode} and \eqref{eqn:sparseMode}, all min candidates with invalid positions in $D$ (i.e., $D[i,j]$ with $i<1$ or $j<1$) are simply ignored.}
\begin{align} 
	\label{eqn:denseMode}
	D[i,j] &= \mathrm{min} \left\{
	\begin{aligned}
		& D[i-1,j-1] + 1 \cdot C[i,j], \\
		& D[i-1,j-2] + 1 \cdot C[i,j], \\
		& D[i-2,j-1] + 2 \cdot C[i,j]
	\end{aligned}
	\right\}
\end{align}
If $g_{i-1} \ne 1$ or $g_{i-2} \ne 1$, then we select the sparse matching mode and calculate $D[i,j]$, $i>1$, $j>1$ as:
\begin{align} 
	\label{eqn:sparseMode}
	D[i,j] &= \mathrm{min} \left\{
	\begin{aligned}
		& D[i-1,j- 2 g_{i-1}] + C[i,j],  \\
		& D[i-1,j- 2 g_{i-1} + 1] + C[i,j], \\
		& D[i-1,j- 2 g_{i-1} + 2] + C[i,j], \\
		& \dots \\
		& D[i-1,j-\left\lceil \frac{g_{i-1}}{2} \right\rceil] + C[i,j]
	\end{aligned}
	\right\}
\end{align}
In other words, the sparse matching ignores the features in the gap region but still enforces a maximum time warping factor of $2$, to be consistent with the dense matching mode.  For example, position $D[4,10]$ in Figure~\ref{fig:DualModeExpl} considers the four possible transitions $(1,1)$, $(1,2)$, $(1,3)$, and $(1,4)$, since the horizontal movement may vary between $\frac{g_3}{2} = 1$ and $2 g_3 = 4$.  As each entry $D[i,j]$ is computed, we also determine $B[i,j]$, which simply indicates which of the entries in equation \eqref{eqn:denseMode} or \eqref{eqn:sparseMode} was selected.

The fourth step is to backtrace.  Since we allow for subsequence matches, we identify the endpoint by determining the minimum element in the last row of D, i.e.,~$\operatorname{argmin} D[L,:]$.  We then follow the backpointers in the backtrace matrix $B$ to determine the optimal path.  There are two important things to note here.  First, interpreting the value of $B[i,j]$ requires checking the gap length values $g_{i-1}$ and $g_{i-2}$ to determine which mode (dense or sparse) was used.  Second, the coordinates in our cost matrix indicate the ordinal position among the \textit{selected} orchestra features, so we must map our cost matrix coordinates to the position among the \textit{dense} orchestra features.  


\begin{table}
	\caption{Comparing the alignment accuracy of several approaches on the piano concerto accompaniment benchmark.  Each number indicates the error rate (in \%) at a fixed error tolerance.}
	\centering
	\begin{tabular}{|l|ccccc|}
		\hline
		\multirow{2}{*}{System} & \multicolumn{5}{c|}{Error Tolerance} \\
		& 0.1s & 0.2s & 0.5s & 1.0s & 2.0s\\
		\hline
		Naive Pairwise DTW & 53.9 & 32.2 & 12.9 & 5.1 & 2.7 \\
		Iterative Subtractive Alignment \cite{yang2021aligning} & 66.5 & 48.0 & 24.3 & 12.3 & 7.5 \\
		Separation + DTW (Spleeter) \cite{ozer2022source} & 64.8 & 46.1 & 23.6 & 11.5 & 4.8 \\
		Separation + DTW (HDemucs) \cite{OezerM24_PianoSourceSep_TASLP} & 55.4 & 33.0 & 10.7 & \textbf{2.1} & \textbf{0.3} \\
		DS-DTW & \textbf{53.1} & \textbf{29.9} & \textbf{10.0} & 2.9 & 0.6 \\
		\hline
	\end{tabular}
	\label{tab:results}
\end{table}

\section{Results}
\label{sec:results}

We compare the performance of five different systems on our benchmark.  The first system is a naive pairwise DTW approach.  We estimate the $P_\mathrm{query}$--$M_\mathrm{ref}$ and $O_\mathrm{acc}$--$M_\mathrm{ref}$ alignments using subsequence DTW (with the same settings described in Section~\ref{subsec:overview}).  We then use these alignments to infer the $P_\mathrm{query}$--$O_\mathrm{acc}$ alignment.  The second system is iterative subtractive alignment \cite{yang2021aligning}, which is based on spectral subtraction.  Here, the mix recording is aligned with the piano-only recording, the estimated alignment is used to perform spectral subtraction of the piano part from the mix, and the modified spectrogram of the mix is then aligned with the orchestra-only recording.  The third and fourth systems are based on source separation (``Separation + DTW'').  Here, source separation is performed on $M_\mathrm{ref}$ to estimate the piano component $\hat{P}_\mathrm{ref}$ and orchestra component $\hat{O}_\mathrm{ref}$.  We then estimate the $P_\mathrm{query}$--$\hat{P}_\mathrm{ref}$ and $O_\mathrm{acc}$--$\hat{O}_\mathrm{ref}$ alignments using subsequence DTW, and use these to infer the $P_\mathrm{query}$--$O_\mathrm{acc}$ alignment.  We experiment with two different pretrained piano-orchestra source separation models based on the U-Net architecture: one that models the spectrogram only (``Spleeter'') \cite{ozer2022source} and another that models both spectrogram and raw waveform (``HDemucs'') \cite{OezerM24_PianoSourceSep_TASLP}.  The fifth system is our proposed DS-DTW algorithm.  This approach is identical to naive pairwise DTW but estimates the $M_\mathrm{ref}$--$O_\mathrm{acc}$ alignment using DS-DTW with $\gamma=0.8$.


Table~\ref{tab:results} compares the alignment accuracy of all systems on our benchmark.  There are three things to notice about these results.  First, the effectiveness of a ``source separation then align'' approach depends heavily on the quality of the source separation.  We can see that source separation with Spleeter yields much worse results than naive pairwise DTW, whereas source separation with HDemucs leads to much better results than naive pairwise DTW at coarser error tolerances.  In a similar vein, the ISA approach performs much worse than naive pairwise DTW.  Second, the DS-DTW approach has consistently better performance than naive pairwise DTW, especially at coarser error tolerances.  For example, it reduces the error rate from $2.7\%$ to $0.6\%$ at 2 sec tolerance ($78\%$ reduction in error rate) and from $5.1\%$ to $2.9\%$ at 1 sec tolerance ($43\%$ reduction).  Third, the approach based on DS-DTW has the best performance at low error tolerances and comparable results to HDemucs at coarser error tolerances.  The main benefit of DS-DTW is its competitive performance with much less complexity than a source separation approach.  Whereas HDemucs requires training a specialized piano-orchestra separation model, DS-DTW has no trainable parameters (only one hyperparameter) and makes no hard assumptions about the instrumentation (piano vs.~violin concerto).

\begin{table}
	\caption{Characterizing the effect of the sparsity hyperparameter $\gamma$ on the performance of DS-DTW.  Each number indicates error rate (\%) at a fixed error tolerance.}
	\centering
	\begin{tabular}{|c|ccccc|}
		\hline
		\multirow{2}{*}{$\gamma$} & \multicolumn{5}{c|}{Error Tolerance} \\
		& 0.1s & 0.2s & 0.5s & 1.0s & 2.0s\\
		\hline
		0.50 & 59.6 & 38.2 & 16.5 & 8.0 & 4.1 \\
		0.60 & 56.0 & 32.6 & 11.6 & 5.4 & 2.3 \\
		0.70 & 54.0 & 31.7 & 11.2 & 4.3 & 2.4 \\
		0.75 & 53.1 & 30.5 & 11.1 & 3.9 & 1.5 \\
		0.80 & 53.1 & 29.8 & 10.0 & 2.9 & 0.6 \\
		0.85 & 53.2 & 31.5 & 11.7 & 4.4 & 2.0 \\
		0.90 & 53.8 & 32.2 & 13.1 & 6.0 & 2.9 \\
		1.00 & 53.9 & 32.2 & 12.9 & 5.1 & 2.7 \\
		\hline
	\end{tabular}
	\label{tab:gamma}
\end{table}

Table \ref{tab:gamma} compares performance of the DS-DTW approach across a range of $\gamma$ settings.  Note that standard DTW is a special case of DS-DTW with $\gamma = 1.0$.  There are two things to notice about these results.  First, we can see that DS-DTW outperforms standard DTW ($\gamma = 1.0$) across a range of $\gamma$ settings, roughly between $0.6$ and $1.0$ for most error tolerances.  This indicates that standard DTW -- considered as a special case of DS-DTW -- is not the optimal setting in many situations.  Second, we can see that tuning $\gamma$ can lead to substantial improvements at coarser error tolerances.  For example, the error rate with a 2 second error tolerance can be reduced from $2.7\%$ to $0.6\%$ ($78\%$ reduction).  This suggests that DS-DTW primarily improves robustness to more ``global'' alignment issues.  Intuitively, the optimal $\gamma$ will depend on characteristics of the audio.  In future work, we plan to explore ways to adaptively select $\gamma$ at test-time.

\section*{Acknowledgment}

This material is based upon work supported by the National Science Foundation under Grant No. 2144050.  This work was also funded by the Deutsche Forschungsgemeinschaft (DFG, German Research Foundation) under Grant No. 500643750 (MU 2686/15-1) and Grant No. 328416299 (MU 2686/10-2). The International Audio Laboratories Erlangen are a joint institution of the Friedrich-Alexander-Universität Erlangen-Nürnberg (FAU) and Fraunhofer Institute for Integrated Circuits IIS.  

\bibliographystyle{IEEEtran}
\bibliography{DSDTW_icassp2025}

\end{document}